\documentclass[12pt,preprint]{aastex}
\usepackage{mkfig}

\begin{document}

\title{\bf Revisiting {\em Ulysses} Observations of Interstellar Helium}

\author{Brian E.\ Wood\altaffilmark{1}, Hans-Reinhard
  M\"{u}ller\altaffilmark{2}, Manfred Witte\altaffilmark{3}}


\altaffiltext{1}{Naval Research Laboratory, Space Science Division,
  Washington, DC 20375, USA; brian.wood@nrl.navy.mil}
\altaffiltext{2}{Department of Physics and Astronomy, Dartmouth College,
  Hanover, NH 03755, USA}
\altaffiltext{3}{Max-Planck-Institute for Solar System Research,
  Katlenburg-Lindau, Germany}

\begin{abstract}

     We report the results of a comprehensive reanalysis of {\em Ulysses}
observations of interstellar He atoms flowing through the solar
system, the goal being to reassess the interstellar He flow vector and
to search for evidence of variability in this vector.  We find no
evidence that the He beam seen by {\em Ulysses} changes at all from
1994--2007.  The direction of flow changes by no more than $\sim 0.3^{\circ}$
and the speed by no more than $\sim 0.3$~km~s$^{-1}$.  A
global fit to all acceptable He beam maps from 1994--2007 yields the
following He flow parameters: $V_{ISM}=26.08\pm 0.21$ km~s$^{-1}$,
$\lambda=75.54\pm 0.19^{\circ}$, $\beta=-5.44\pm 0.24^{\circ}$, and
$T=7260\pm 270$~K; where $\lambda$ and $\beta$ are the ecliptic
longitude and latitude direction in J2000 coordinates.  The flow
vector is consistent with the original analysis of the
{\em Ulysses} team, but our temperature is significantly higher.  The
higher temperature somewhat mitigates a discrepancy that exists in the
He flow parameters measured by {\em Ulysses} and the {\em Interstellar
Boundary Explorer}, but does not resolve it entirely.  Using
a novel technique to infer photoionization loss rates directly from
{\em Ulysses} data, we estimate a density of
$n_{\rm He}=0.0196\pm 0.0033$ cm$^{-3}$ in the interstellar medium.

\end{abstract}


\section{INTRODUCTION}

     The global structure of the heliosphere is determined
in large part by the local ISM flow vector in the rest frame of the
Sun.  Older determinations of this vector, particularly those based on
measurements of interstellar He flowing through the inner heliosphere
by the GAS instrument on {\em Ulysses}, have recently been challenged
by new measurements of the He flow by the {\em Interstellar Boundary
Explorer} (IBEX).  The canonical {\em Ulysses} analysis of \citet{mw04}
suggested the following He flow parameters: $V_{ISM}=26.3\pm
0.4$ km~s$^{-1}$, $\lambda=75.4\pm 0.5^{\circ}$, $\beta=-5.2\pm
0.2^{\circ}$, and $T=6300\pm 340$~K; where $\lambda$ and $\beta$ are
the ecliptic longitude and latitude direction in J2000 coordinates.
In contrast, IBEX data recently yielded the following flow properties
\citep{mb12,em12,djm12}:
$V_{ISM}=23.2\pm 0.3$ km~s$^{-1}$, $\lambda=79.00\pm 0.47^{\circ}$,
$\beta=-4.98\pm 0.21^{\circ}$, and $T=6300\pm 390$~K.  Significant
inconsistencies exist between the {\em Ulysses} and IBEX measurements,
particularly for $V_{ISM}$ and $\lambda$.

     The lower $V_{ISM}$ value of IBEX is of particular interest,
as it may imply the nonexistence of a bow shock around the
heliosphere.  The $V_{ISM}=23-27$ km~s$^{-1}$ interstellar flow speed
happens to imply a fast magnetosonic Mach number of $M_F\approx 1$, making
the existence or nonexistence of a bow shock around the heliosphere
very much an open question.  Uncertainties in the strength and
orientation of the interstellar magnetic field, $B_{ISM}$, represent
one obstacle in determining whether a bow shock exists.  The higher
$B_{ISM}$ is (and the more perpendicular to the ISM flow), the lower
$M_F$ should be, and the less likely there is to be a bow shock.  But
uncertainties in $V_{ISM}$ are also important.

     For many years the best assessments of ISM velocity suggested
$V_{ISM}\approx 26$ km~s$^{-1}$, not only the {\em Ulysses}/GAS
measurements \citep{mw96,mw04} but measurements from
ISM absorption lines as well ($V_{ISM}=25.7\pm 0.5$ km~s$^{-1}$;
Lallement \& Bertin 1992).  With these relatively high values,
heliospheric modelers favored $M_F > 1$, implying the existence of a
bow shock.  However, not only has the new IBEX measurement called this
into question, but so has one reanalysis of ISM absorption line data
\citep{sr08}, which suggests $V_{ISM}=23.84\pm 0.90$
km~s$^{-1}$.  The lower $V_{ISM}=23-24$ km~s$^{-1}$ values reported
recently have been enough for many to argue that $M_F < 1$ should be
preferred \citep{djm12,bz13}.  However,
\citet{ks14} argue that including He$^+$ density in the
calculation of Alfv\'{e}n speeds instead of just assuming a
pure proton plasma would still suggest $M_F > 1$ even if
$V_{ISM}\approx 23$ km~s$^{-1}$.  On the other hand, \citet{gpz13}
note that charge exchange processes may turn the bow
shock into more of a ``bow wave'' even if $M_F > 1$.

     In any case, the existence or nonexistence of a heliospheric bow
shock is one issue driving interest in whether the He flow vector
of {\em Ulysses} or that of IBEX is to be preferred.  Assuming
that the ISM cloud around the Sun is nonrigid, \citet{cg14} have
reanalyzed ISM absorption line data and infer cloud kinematics
more consistent with the {\em Ulysses}
velocity being valid close to the Sun.
\citet{lbj13}, \citet{rl14}, and \citet{fev14} have
recently made additional arguments in favor of the older {\em Ulysses}
vector, implying that the IBEX measurements must somehow be in error.
\citet{pcf13} propose a very different solution.  They
propose that the local ISM flow actually varies with time, and
that these variations are responsible for the {\em Ulysses}/IBEX
discrepancy.  However, observations of interstellar H from the
Solar Wind Anisotropies instrument on the {\em Solar and Heliospheric
Observatory} and the Space Telescope Imaging Spectrograph instrument on
the {\em Hubble Space Telescope} do not provide support for a variable
ISM flow vector \citep{fev14}.

     All this attention on the He flow vector provides motivation
for taking another long look at the old {\em Ulysses}/GAS data, in order
to see if a new, independent analysis confirms the \citet{mw04} results, and
to see if there is any evidence for He flow variability in the
{\em Ulysses} data that would support the \citet{pcf13}
interpretation favoring a variable ISM flow vector.  Further justification
is that while the \citet{mw04} analysis considered data taken from
1990--2002, later data acquired up until the end of the {\em Ulysses}
mission in 2007 were not considered.

     \citet{oak14} have recently analyzed a couple of
{\em Ulysses} He beam maps, including one from 2007,
demonstrating that the new IBEX He vector cannot reproduce either
map.  They also note that the \citet{mw04} vector might in principle
be able to fit the IBEX data if the temperature is allowed to be $\sim
9000$~K, consistent with more extensive IBEX analyses \citep{mb12,em12,djm12}.
We here present
our own comprehensive reanalysis of the full {\em Ulysses}/GAS data
set, providing a new assessment of the He flow vector using analysis
techniques developed independently from both the original
{\em Ulysses}/GAS team \citep{mw96} and the IBEX team
\citep{mb12,em12}.

\section{THE {\em ULYSSES} DATA}

    Launched in 1990 October, the primary mission of the {\em Ulysses}
spacecraft was to study various interplanetary constituents like the
solar wind, magnetic fields, radio waves, dust, solar and galactic
cosmic rays, etc.\ outside the ecliptic plane in an orbit passing over
the solar poles \citep{kpw92,rgm01}.  This orbit was
accomplished with a gravitational
assist from Jupiter in 1992 February, leading to an orbit nearly
perpendicular to the ecliptic plane, with an aphelion near Jupiter's
distance of 5~AU, and a perihelion near 1~AU.  {\em Ulysses} continued
to make observations from this orbit until the mission ended in late
2007.  The {\em Ulysses}/GAS instrument, described in detail by \citet{mw92},
was designed primarily to study interstellar neutral
He atoms flowing through the inner heliosphere.  Neutral helium
provides the best diagnostic of the undisturbed ISM flow due to the
high abundance of helium, and due to the fact that helium is not
greatly affected by charge exchange during its journey through the
outer heliosphere, in contrast to hydrogen and oxygen, for example.

     The first He observations were made by {\em Ulysses}/GAS during
its cruise phase out to Jupiter (1990--1992).  These data were first
studied by \citet{mw93}.  However, in this paper we will focus
on data taken during the operational phase of the {\em Ulysses}
mission, with {\em Ulysses} in its polar orbit after the Jupiter
encounter.  These He data are generally of higher quality than the
cruise phase data \citep{mw04}, with better sampling of the He beam.
Additional advantages of the operations phase data will become very
clear below.

     Successful detection of interstellar He required that the
inflow velocity be high enough for the He atoms to exceed the particle
detection threshold of the GAS instrument.  With the {\em Ulysses}
orbital plane being basically perpendicular to the ISM flow vector,
the detectability of He depended mostly on the spacecraft velocity.
Only when {\em Ulysses} was in the near-Sun part of its orbit was the
spacecraft traveling rapidly enough so that the incoming He had
sufficient energy to exceed the detection threshold.  Thus, the
{\em Ulysses}/GAS He observations were confined to the three fast
latitude scans made by {\em Ulysses}, in the periods 1994--1996,
2000--2002, and 2006--2007.  These are the three epochs that we will
focus on here.  \citet{mw04} considered the first two epochs of
observations, but not the last epoch.

     The GAS device works essentially like a pin-hole camera with
a single-element detector,
so observing the He beam requires making many different exposures in
different directions to fully map out the count rate distribution.
The {\em Ulysses} spin axis, which was always pointed towards Earth,
defined how this was done.  {\em Ulysses} had a rotation period of
12~s.  During this period, the GAS instrument would observe at a
particular elevation angle, measured relative to the spin axis,
mapping out a ring on the celestial sphere.  Detected counts were
accumulated in 32 bins, which could uniformly cover the full
$360^{\circ}$ rotation; or more commonly cover smaller arcs of
$90^{\circ}$, $45^{\circ}$, or $22.5^{\circ}$.  After a period of
typically 68~min, the elevation angle scanned by GAS would be changed
by a step of $1^{\circ}$, $2^{\circ}$, $4^{\circ}$, or $8^{\circ}$;
allowing another arc to be observed.  Over the course of $2-3$ days,
typically, a full scan across the He beam would be completed, and a
new map begun.

     The GAS instrument actually utilized two different channels with
independent detectors, one with a wide field of view (WFOV) and one
with a narrow field of view (NFOV), and both channels were exposed
simultaneously.  Over the course of the three fast latitude scans
mentioned above, {\em Ulysses} obtained $\sim 400$ WFOV maps and $\sim
400$ NFOV maps.  We will be working almost exclusively with the WFOV
data here, which have higher signal-to-noise (S/N) than the NFOV maps.

     Figure~1 shows a typical WFOV map, from 2001~January~24.  The GAS
instrument suffers from some sensitivity to UV emission even when in
its particle detection mode \citep{mw96}, so some hot
stars are visible in the map, in addition to the broad He beam.  The
stars can be inconvenient if blended with the He beam, but are
actually useful as calibrators of pointing accuracy otherwise.  In
this article we will in fact be using a recent recalibration of the
{\em Ulysses}/GAS data set making extensive use of the stars as
pointing calibrators, which is described in more detail by \citet{mb14}.
In an end-to-end test, the calibration of the direction
determination was adjusted so that the positions of stars observed
by the instrument coincided with their astronomical position.
Based on many more star observations from the whole mission a minor
further adjustment of a few tenth of a degree was required and
subsequently all He data were reprocessed in 2013 with this new
calibration.  This recalibration provides yet another motivation for
reanalyzing these data, though the change in the data is modest.
Note that although the archival {\em Ulysses} data, and its original
analyses reported in the Witte et al.\ series of papers, are in
B1950 coordinates, the
numbers and figures we provide in this article will be in the more
modern J2000 system.

     A final distinction to be made is that we will only be
considering observations of the so-called ``direct beam'' formed by He
atoms that flow directly to the detector (with modest gravitational
deflection), as opposed to the ``indirect beam'' created by particles
that go around the Sun and are gravitationally redirected back towards
the detector, arriving from a very different direction than the direct
beam atoms.  The indirect beam is much fainter than the direct beam
due to massive photoionization and electron-impact ionization losses
that occur when these particles pass close to the Sun.  {\em Ulysses}
has nevertheless detected the indirect beam in observations from 1995
\citep{mw04}, but we will not be considering those data here.

\section{EMPIRICAL ANALYSIS}

     Rather than immediately try to extract a He flow vector from
the data, which requires a somewhat complex forward modeling analysis
(see Sections 4 and 5),
we find it worthwhile to first make some purely
empirical investigations of the beam.  The primary motivation
for this is to look for any evidence that the beam is changing at
all during the 1994-2007 time period.  We quantify the properties of
the beam using 2-D Gaussian fits to the beam, as mapped into ecliptic
coordinates (as in Figure~1).  Note that there is
no physical reason the beam should be precisely Gaussian,
but in practice the beams observed by {\em Ulysses} are reasonably
Gaussian in shape, so this is a practical way to quantify
them.  An example Gaussian fit is shown in Figure~2.  There
are six free parameters of such a fit:  the central longitude and
latitude, the latitudinal and longitudinal widths [specifically
the full-width-at-half-maximum (FWHM)], the count rate amplitude at
beam center, and a background count rate; where we assume a flat
background underneath the He beam.

     In the following discussion, we make use of an orbital phase
for {\em Ulysses}, $\phi$, based on the angle defined by the positions
of the spacecraft and the Sun, and the location of the ecliptic plane
crossing near aphelion.  In this system, the ecliptic plane crossing
near aphelion at ecliptic coordinates
($\lambda$,$\beta$)=($157.4^{\circ}$,$0^{\circ}$) is at $\phi=0.0$, and that
near perihelion at ($337.4^{\circ}$,$0^{\circ}$) is at $\phi=0.5$.
The $\phi=0.25$ and $\phi=0.75$ orbital phases correspond to the
near south pole and near north pole locations (relative to the Sun),
so {\em Ulysses} is below the ecliptic at $\phi<0.5$ and above it at
$\phi>0.5$.  ({\em Ulysses} travels from south to north during its
fast latitude scans.)

     The Gaussian fits are performed using a semi-automated procedure,
but all the fits are visually verified.  Many beam maps end up being
excluded for various reasons.  At $\phi<0.2$ and $\phi>0.9$ the data
are simply deemed too noisy.  Observations with $\phi<0.26$ are excluded
due to contamination from a star (specifically the rightmost star seen
in Figure~1).  Some maps are excluded for having a background that is
not flat, possibly due to solar activity at that time.  The most
common reason for exclusion is simply that some scans did not fully
cover the He beam.  We only want to consider beam maps where the scan
fully encompasses the He beam.  Ultimately, fits are accepted for 234
NFOV maps and 238 WFOV maps, including the map in Figure~1.  As
mentioned above, we will be focusing on the higher S/N WFOV data, but
we always do verify that results are not different when considering
the NFOV data instead.

     The Gaussian fit parameters of the WFOV maps are plotted in
Figures~3 and 4.  Figure~3a shows that the He beam traces out a horseshoe
pattern in the sky during a fast latitude scan.  Performing polynomial
fits to longitude and latitude as a function of orbital phase allows
us to define the average beam location illustrated in Figure~3a.
We will emphasize below how the shape of this horseshoe is a powerful
diagnostic for the He flow parameters.  Figure~3a demonstrates that
basically the same horseshoe pattern is made during all three
fast latitude scans, showing little indication for variation in
beam location.  Figures~3b-d provide a better indication for the
deviations of beam position about the average.  The scatter
of the data points is indicative of the measurement uncertainties, but
some systematic deviations are discernible despite the scatter.  For
example, the beam longitude appears to be lower during epoch 3 by
a few tenths of a degree.  The beam latitude is also a few tenths of
a degree lower during epoch 3, but only for $\phi=0.5-0.7$.  The
latitudes observed during epochs 1 and 2 are discrepant for
$\phi=0.75-0.85$.

     Of particular interest are longitude variations, as it is in
longitude that the IBEX and {\em Ulysses} flow vectors are in
disagreement, and it is in longitude that \citet{pcf13} claim evidence
for a systematic increase with time.  There is, however, no evidence
for any increase in longitude during the {\em Ulysses} era, with the
epoch 3 data suggesting if anything a slight {\em decrease}, as
noted above.  The average discrepancy in Figure~3d is
$0.28^{\circ}$.  This can be interpreted as an estimate of
the pointing uncertainties inherent in the {\em Ulysses} data,
consistent with expectations based on the pointing calibration done
using stars \citep{mw04,mb14}.

     In any case, the longitude variations in Figure~3b are much
smaller than the $3.6^{\circ}$ discrepancy in longitude between the
IBEX and {\em Ulysses} He flow vectors.  We conclude that there is no
indication from the {\em Ulysses} data to support the notion that He
flow variability is responsible for the IBEX/{\em Ulysses}
discrepancy.  Furthermore, given that we now know that the 2007
{\em Ulysses} data are not more consistent with the IBEX results than
the older data, there is only a 2 year time gap between the
{\em Ulysses} and IBEX eras for the He flow to have changed so
significantly.

     Figures~4c-d also demonstrate that there is no evidence
for any significant change in the He beam shape from 1994--2007.
The latitudinal and longitudinal beam widths change with orbital
phase, but the same sort of variation is observed during all three
fast latitude scans.  Figures~4a-b show that the three epochs do,
however, exhibit noticeable differences in background and beam
amplitude.  The {\em Ulysses}/GAS background is a complicated mix of
solar and cosmic sources, and we will not discuss it much here, other
than to note that both orbital phase and epochal variations are
indicated by Figure~4b.

     As for the beam count rate amplitude in Figure~4a, it naturally
peaks near $\phi=0.5$ during each fast latitude scan, as that is when
the spacecraft is moving fastest.  The different count rate levels
observed during the three epochs are most naturally interpreted as
being indicative of variations in the solar EUV photoionization rate.
It makes perfect sense for epoch 2 (2000--2002) to have the lowest
He fluxes, as this is a time near solar maximum, with high
photoionization rates and therefore substantial losses of neutral
He due to photoionization.  In contrast, epochs 1 and 3 are both
near solar minimum, thereby minimizing photoionization losses.
We will return to the related issues of photoionization and
He density, $n_{\rm He}$, in Section~5.


\section{MEASURING THE {\em ULYSSES} He FLOW VECTOR}

     The simple quantification of the He beam properties provided
by the Gaussian fitting in the previous section is not helpful for
actually inferring a He flow vector from He beam maps, which instead
requires a forward modeling approach.  In this procedure, a
Maxwellian He velocity distribution is assumed at infinity,
characterized by five parameters ($n_{\rm He}$, V$_{ISM}$, $\lambda$, $\beta$, T),
and then this Maxwellian is propagated to the position
of the spacecraft.  The distribution will be primarily affected by the
Sun's gravitational influence.  Loss processes can also be accounted
for in this forward modeling, the most important being photoionization
by solar EUV photons.  We will initially ignore photoionization in our
forward modeling, but we will consider it later to see if it
changes our inferred He flow parameters.

     The forward modeling method utilized in the original
{\em Ulysses}/GAS team analyses is described by \citet{mb96}.
Our forward modeling routine, which is described in
detail elsewhere \citep{hrm12b,hrm12a,hrm13},
involves a rather different approach to the
problem than the usual method involving numerical integration from
infinity.  In our computations, quantities that are conserved along a
trajectory, such as total energy, angular momentum, eccentricity, and
direction of perihelion, are used to determine in one algebraic step
the neutral He velocity at a desired point in space when its velocity
at infinity is given.  Conversely, when both location and velocity of
a He atom are given, the conservation equations dictate a unique
corresponding velocity at infinity.  The entire phase space
distribution, F({\bf v}), can be calculated at a desired point in the
inner heliosphere by decomposing the problem into a family of
trajectories with different velocities (different conserved
quantities) that all converge at the chosen point.  By assuming a
thermal Maxwellian as the original distribution at infinity, each of
those trajectories can then be assigned its corresponding phase space
density, yielding F({\bf v}) at the point of interest
\citep{hrm12b,hrm12a,hrm13}.

     Computing count rates in a given direction
involves first transforming F({\bf v}) into the spacecraft
rest frame and converting into a coordinate system defined around the
view direction of interest, resulting in F($V_s$,$\theta$,$\psi$),
where $V_s$ is the speed relative to the spacecraft, $\theta$ is
the poloidal angle with respect to the view direction, and $\psi$ is the
azimuthal angle.  With a volume element
``$d^3v=V_s^2\sin{\theta}~dV_s~d\psi~d\theta$'', the predicted count rate
can then be expressed as
\begin{equation}
C=A \int V_s~D_e(V_s)~G(\theta)~F(V_s,\theta,\psi)~d^3v,
\end{equation}
where $A=0.0908$~cm$^2$ is the effective area, $D_e(V_s)$ is the
detection efficiency, and $G(\theta)$ is the detector geometry
function.  Figure~5 shows our digitized versions of $G(\theta)$
and $D_e$, approximated from Figures~1 and 3 of \citet{mb96},
with the latter expressed as a function of incoming
particle energy, $E_s=\frac{1}{2}mV_s^2$, with $m$ being the mass of
the He atom.  For a given elevation angle, particles in the
azimuthal bins described in Section~2 are observed while the
spacecraft is spinning, so there is ``scan smoothing'' along the
azimuthal direction.  We account for this by averaging $C$ values
computed at pixel center and at $\pm \Delta \alpha/2$,
where $\Delta \alpha$ is the azimuthal angular bin size.

     By computing $C$ for all $n$ observed directions,
we compute a synthetic He beam map, $C_i$ (for i=1,n), for
comparison with the real one.  This
synthetic beam is multiplied by a normalization factor, $N$, in
order to best match the count rates of the observed beam.  This
amounts to a revision of the assumed $n_{\rm He}$ value to
$n^{\prime}_{\rm He} = N\times n_{\rm He}$.  In
making this normalization we are in effect replacing $n_{\rm He}$
with $N$ as the true density-related free parameter of our fit.
We also have to account for the background count
rate, $B$.  As we did in the Gaussian fitting, we simply assume a flat
background.  Thus, the new predicted count rates are
\begin{equation}
C^{\prime}_i=N\times C_i + B.
\end{equation}
We perform a $\chi^2$ minimization \citep[e.g.,][]{prb92}
to determine the $N$ and $B$ values that lead to the best fit to the
data.  The $\chi^2$ value of this best fit represents our measure of
merit for this particular synthetic He beam map, which can be
expressed as
\begin{equation}
\chi^2=\sum_{i=1}^{n} \left( \frac{S_i-C^{\prime}_i}{\sigma_i} \right)^2,
\end{equation}
where $C^{\prime}_i=C^{\prime}_i[N,B,C_i(V_{ISM},\lambda,\beta,T)]$ are
the predicted count rates, $S_i$ are
the observed count rates, $\sigma_i$ their Poissonian uncertainties, and
$n$ the number of points considered in the He beam map.
We then have to determine what
He flow parameters ($V_{ISM}$, $\lambda$, $\beta$, and T) truly minimize
$\chi^2$, representing the best fit to the data.

     The standard approach for determining a set of best-fit parameters
is to start with a guess for the parameters, compute $\chi^2$, and
then use a $\chi^2$ minimization routine to refine the parameters to
reduce $\chi^2$ until ultimately the true $\chi^2$ minimum is reached.
If $\nu$ is the number of degrees of freedom of the fit (the number of
data points minus the number of free parameters), then the reduced
chi-squared is defined as $\chi^2_{\nu}=\chi^2/\nu$, which should be
$\sim 1$ for a good fit.  We use the Marquardt method described by
\citet{whp89} as our $\chi^2$ minimization routine.

     Figure~6a shows a WFOV map from 1995~August~20, and Figure~6b-c
shows a fit to this map, with the \citet{mw04} vector used as our
initial guess for the fit parameters.  It is a simple matter for
the particle tracking code mentioned above \citep{hrm12b,hrm12a}
to compute the expected beam location based on the \citet{mw04}
vector, if we assume $T=0$~K (meaning all He atoms have
the exact same velocity and follow the exact same trajectory).  We use
this prediction to define the map region to be fitted.  In particular,
we only consider data points within $25^{\circ}$ of this location
(white dots in the figure), which encompasses enough of the region
beyond the He beam to estimate the background.
Figure~6b shows the synthetic beam map of the best fit, and Figure~6c
explicitly plots observed versus predicted count rates, with a data
point plotted for each of the white dots in Figure~6b.  The He flow
vector parameters listed in Figure~6c are gratifyingly close to the
\citet{mw04} vector, and the $\chi^2_{\nu}=1.06934$ value indicates an
excellent quality fit.

     Unfortunately, the best-fit parameters that we infer are actually
highly dependent on our initial guesses for the parameters.  This is
shown explicitly in a second fit to the 1995~August~20 data in
Figure~6d-e.  With an initial guess very different from the \citet{mw04}
vector used in Figure~6b-c, we end up with a fit to the data
with very different parameters.  The parameters listed in
Figure~6e are very far from all modern estimates of the local ISM
flow vector, so we know they are wrong.  But these parameters lead
to a fit that is just as good as the more plausible fit in Figure~6b,
with a $\chi^2_{\nu}=1.05731$ value that is actually slightly lower even.

     This exercise emphasizes just how difficult it is to find a
unique He flow vector from a single He beam map.  A single map does
not sufficiently constrain the problem.  There are significant
degeneracies among the fit parameters, leading to long diagonal
troughs in $\chi^2$ space where it is very hard to find the true
$\chi^2$ minimum along the trough.  This is true for the analysis of
IBEX data as well \citep[see Figure~22 in][]{mb12}.  The most
basic degeneracy is between flow direction and flow speed, which can
be explained by the following question: If a beam is seen from a given
direction, is that due to a fast beam coming from roughly that
direction, or due to a slower beam that has been gravitationally
deflected into that direction?  Since the He flow vector happens to
lie near the ecliptic plane, this degeneracy basically ends up mostly
between $V_{ISM}$ and $\lambda$ (rather than $\beta$).

     We argue here that these degeneracies can be effectively
broken if {\em Ulysses} data from different parts of the {\em Ulysses}
orbit are fitted together, as opposed to fitting one beam map at a
time, as in Figure~6.  The utility of this approach is
illustrated by Figure~7.  We had previously emphasized how the He beam
traces out a horseshoe pattern in the sky during a fast latitude scan
(see Figure~3a).  Figure~7 shows the horseshoe pattern expected based
on the \citet{mw04} vector (assuming $T=0$~K), and also shows what
happens to the horseshoe when $\beta$, $\lambda$, and $V_{ISM}$ are
changed.  When $\beta$ and $\lambda$ are changed the horseshoe shifts
vertically and horizontally, respectively.  In contrast, $V_{ISM}$
affects the size of the horseshoe, with higher $V_{ISM}$ yielding a
smaller horseshoe.  The crucial point is that $V_{ISM}$ affects the
horseshoe shape very differently from $\beta$ and $\lambda$, thereby
breaking the direction/velocity degeneracy noted above.  But this
diagnostic power can only be properly considered if He beam maps from
all parts of the horseshoe are considered simultaneously in fitting
the data; hence the global fit approach.

     The specific data used in this global fit approach are the 238
WFOV maps used in the Gaussian fitting (see Section~3).  The global
fit is similar to the individual map fits in Figure~6, but we are now
fitting many more data points, naturally; 81,405 count rates within
the 238 maps, to be precise.  In addition to the four He flow parameters
of interest ($V_{ISM}$, $\lambda$, $\beta$, and $T$), there are also
$N$ and $B$ parameters for each individual map, for a total of
480 free parameters.  Summing over the 238 separate maps,
equation (3) becomes
\begin{equation}
\chi^2=\sum_{j=1}^{238} \sum_{i=1}^{n} \left( \frac{S_{j,i}-
  C^{\prime}_{j,i}}{\sigma_{j,i}} \right)^2,
\end{equation}
where $C^{\prime}_{j,i}=C^{\prime}_{j,i}[N_j,B_j,C_{j,i}(V_{ISM},\lambda,\beta,T)]$.
Figure~8 shows the 238 $N_j$ and $B_j$ values that result from our best
fit.  The $B_j$ values are essentially identical to the backgrounds
shown in Figure~4b, measured in the Gaussian fit analysis.
The $N_j$ factors are directly related to inferred He densities (see
above), so that is how they are shown in the figure.  The clear phase
dependence is an effect of photoionization losses, which are more
severe at intermediate phases when the spacecraft is closer to the
Sun.  We will discuss this issue in detail in Section~5 when we
actually seek to correct for photoionization.

     We need to not only find the best fit but also to
estimate uncertainties in the fit parameters, so the full analysis
actually involves a series of fits with one of the four flow
parameters held constant, as we scan through the parameters to see how
$\chi^2$ varies with $V_{ISM}$, $\lambda$, $\beta$, and $T$.  Results
are shown in Figure~9.  There is a very clear $\chi^2$ minimum,
$\chi^2_{min}$, in each panel.  We define $\Delta \chi^2 \equiv
\chi^2-\chi^2_{min}$, and each panel of Figure~9 shows the variation
of $\Delta \chi^2$ across the $\chi^2_{min}$ region.  Fifth order
polynomials are fitted to the data points to interpolate between them.
The $\Delta \chi^2$ values are used to define the error bounds around
$\chi^2_{min}$.  \citet{prb92} and \citet{whp89}
both provide useful discussions about how best to do this.  For the
number of free parameters of our fit (480), the $3\sigma$ confidence
contour corresponds to $\Delta \chi^2=578.3$, based on relation
26.4.14 of \citet{ma65}.  This is the $\Delta \chi^2$
level used to define the uncertainty ranges shown in Figure~9.

     Our final derived He flow parameters are: $V_{ISM}=26.08\pm 0.21$
km~s$^{-1}$, $\lambda=75.54\pm 0.19$~deg, $\beta=-5.44\pm 0.24$~deg,
and $T=7260\pm 270$~K; which are listed in the first line of Table~1.
Our best global fit has $\chi^2_{\nu}=1.524$, which is somewhat
higher than it should be ideally, indicating a modest level of
systematic discrepancy from the data.  It is not practical to show the
actual fit to the data for all 238 beam maps, but for illustrative
purposes Figure~10 shows the fit for two particular maps, one from
1994~November~9 and another from 2007~November~26.  The former
exhibits no clear systematic deviation between the fit and the data, but
the latter shows some systematic deviation.

     Calculations involving photoionization will be discussed in
more detail in Section~5, but at this point we mention that in order
to see whether photoionization could be affecting our He vector and T
measurements, we tried redoing the global fit assuming a He
photoionization rate at 1~AU of $\beta_{ion}=1.5\times 10^{-7}$
s$^{-1}$, which is near the high end of those typically estimated
\citep{mb12,oak14}.  The resulting parameters and $\chi^2$ are:
$V_{ISM}=26.11\pm 0.21$ km~s$^{-1}$, $\lambda=75.47\pm 0.18$~deg,
$\beta=-5.49\pm 0.23$~deg, $T=7330\pm 280$~K,
$\chi^2_{\nu}=1.528$.  These values are very similar to those found
before, so considering photoionization does not change the
derived vector or temperature significantly, and does not affect the
quality of fit, consistent with the findings of \citet{oak14}.

     Figure~9 and Table~1 demonstrate that our results are generally
in very good agreement with the previous canonical {\em Ulysses}
measurements from \citet{mw04}, with the exception of $T$.  Our
temperature is significantly higher than the $T=6300\pm 340$~K
measurement of \citet{mw04}.  As discussed in Section~2, we are
working with a new reduction of the {\em Ulysses} data, but we have
performed fits to He beam maps from the older data reduction, and we
do not find any significant effect on our derived temperature.  We initially
suspected that our higher $T$ measurement may be an artifact of the
global fitting approach.  To first order, $T$ is determined mostly by
the size of the observed He beams, while the other three parameters
determine its direction.  If the global fit has difficulty
simultaneously reproducing the beam center for all 238 beam maps, it
will try to compromise by broadening the beam, increasing the inferred
$T$.  This effect will not be an issue with single-map fits, and so
would not affect the \citet{mw04} analysis approach.

     We perform single-map fits to our 238 beam maps, where we fix
$V_{ISM}=26.08$ km~s$^{-1}$ in order to resolve the parameter degeneracy
problem with single-map fits described above.  This provides 238
measurements of $\lambda$, $\beta$, and $T$.  The average and standard
deviation of these values are reported in the second line of Table~1.
The $\lambda$ and $\beta$ measurements are nearly identical to the
global fit values.  The $T=7090\pm 370$~K measurement is somewhat
lower than the $T=7260\pm 270$~K global fit value, as expected, but
not by that much.  Experimenting with other velocity values within
our quoted $V_{ISM}=26.08\pm 0.21$ km~s$^{-1}$ range does not change
things significantly.  We conclude that the beam-broadening effect of
the global fit is not the dominant cause of our higher $T$.

     Differences in background treatment are another possible cause
of the $T$ discrepancy.  The original {\em Ulysses} data analyses
excluded pixels with count rates too close to the background value
\citep{mb96}.  Typically only the upper $\sim 80$\% of the profile
was considered, whereas our approach is to consider all pixels within
$25^{\circ}$ of the beam, with the background determination actually
being part of the fit rather than assuming a predetermined background.
We have considered the possibility that focusing more on the central
part of the beam could lead to the perception of a narrower beam and
therefore a lower $T$.  However, using single-map fits we have
investigated the effects of focusing on the central part of the beam
and we do not find that such fits consistently lead to significantly
lower temperatures.  Thus, we ultimately dismiss this possible
explanation for our high $T$ measurement.

     We note that our higher temperature is in better agreement with the
$T=7500\pm 1300$~K Local Interstellar Cloud (LIC) temperature
estimated from ISM absorption lines \citep{sr08}.  It is
also consistent with another contemporaneous reassessment of the
{\em Ulysses} He data by \citet{mb14}, largely following
the same analysis approach used previously to study IBEX data \citep{mb12}.
Their derived best-fit He flow parameters ($V_{ISM}=26.0$
km~s$^{-1}$, $\lambda=75.3^{\circ}$, $\beta=-6^{\circ}$, $T=7500$~K)
are in reasonably good agreement with our results, including the
higher temperature.

     The only difference is that \citet{mb14} claim much
larger uncertainties.  Specifically, the parameters are quoted as
having uncertainty ranges of $74.2^{\circ}<\lambda<76.5^{\circ}$,
$-7^{\circ}<\beta<-5^{\circ}$, $24.5<V_{ISM}<27.0$ km~s$^{-1}$, and
$5500<T<9000$~K.  The large error bars are based on independent
analyses of data taken from different parts of the {\em Ulysses}
orbit, but we consider these error bars to be overly conservative,
as this analysis approach greatly weakens the diagnostic power of
considering data from all parts of the {\em Ulysses} orbit
simultaneously, which is crucial to the efficacy of the global fit
approach.  Figure~7 indicates how $\lambda$, $\beta$, and $V_{ISM}$
change the shape of the horseshoe-shaped track for the He beam
in different ways, and it is this effect that allows the global
fit approach to break the degeneracies that plague
single-map analyses (see Figure~6).
Considering observations from only one
part of the {\em Ulysses} orbit will in effect reintroduce the
degeneracy problem, naturally leading to the inference of
the large uncertainties quoted by \citet{mb14}.  To put it another
way, we see the large uncertainties quoted by \citet{mb14} as
being a further demonstration of the importance of considering
data from {\em all} parts of the {\em Ulysses} orbit together,
rather than being an accurate quantification of systematic
uncertainties inherent in these data.

     The cause of the higher T inferred both by ourselves and
\citet{mb14} remains to be explained.  Having excluded the
possibilities mentioned above, we conclude that the most likely cause
of this discrepancy with \citet{mw04} lies in the Bayesian statistics
of the older analyses.  While our analysis and that of \citet{mb14}
rely solely on the $\chi^2$ statistic to quantify how well a set of
parameters is fitting the data, the older analyses instead use a
Bayesian merit function that considers not only consistency with the
data, but also consistency with prior assumptions about what the
best-fit parameters should be \citep[see eqn.\ 10 from][]{mb96}.  The
advantage of this Bayesian approach is that this is one way to break
the degeneracies in single-map fits, and derive relatively
self-consistent single-map measurements like those shown in
Figure~2 of \citet{mw04}.  The disadvantage is that the
analysis is potentially biased by the a~priori assumption of
expected best-fit parameters, as well as the question of how
to weight the relative importance of consistency with those
parameters and consistency with the data.  The clearest distinction
between our analysis and the older ones lies in this difference
in how the best fit is defined, so we propose
that the Bayesian approach of the older work drove the analysis towards
a lower temperature, while the pure $\chi^2$-minimization analysis
used here and by \citet{mb14} leads to a higher temperature.

     The higher temperature suggested by both our analysis and that
of \citet{mb14} could have relevance for the discrepancy in
the IBEX and {\em Ulysses} He flow vectors.  The IBEX analysis suffers
from the same kind of parameter degeneracies discussed above for
{\em Ulysses}.  Although the IBEX data suggest a $\chi^2$ minimum
for $T\approx 6200$~K \citep{mb12}, it is not clear that
significantly higher temperatures are excluded by the data.  Furthermore,
the parameter dependencies for IBEX are such that assuming a higher
temperature would push the IBEX-derived $\lambda$ and $V_{ISM}$ values
closer to those of {\em Ulysses}.  Specifically, assuming our
$T=7260\pm 270$~K measurement would yield the following values
from IBEX:  $\lambda=77.1\pm 0.5^{\circ}$ and
$V_{ISM}=24.6\pm 0.4$ km~s$^{-1}$ \citep{mb12,em12,djm12}.
These values are still
not quite consistent with {\em Ulysses}, but they are significantly
closer than the \citet{djm12} values quoted in Section~1.

     A likely important source of systematic uncertainty in the
{\em Ulysses} data analysis that we want to emphasize is uncertainty
in the detection efficiency ($D_e$) curve in Figure~5b.  It is important
to note that He atoms are typically detected in the $E_s=30-60$~eV range.
This is very much in the ``knee'' of the $D_e$ curve, where $D_e$ is
significantly energy dependent.  Combine this energy dependence with
the fact that there is generally a significant gradient of average
particle energy across a He beam, and the result is a situation where
particle detection efficiency is higher on one side of the beam than
the other, leading to an observed beam that is shifted from where
one might naively expect to see it.  This is clearly the case in
Figure~6 for example, where the observed beam is shifted to the left
from its expected location based on a zero-temperature calculation
assuming the \citet{mw04} vector.

     This effect is shown in more detail in Figure~11, which compares
the observed trace of the He beam center from Figure~3a with a
trace computed from our best-fit He flow vector, assuming $T=0$~K.
The average discrepancy between these two horseshoe-shaped tracks
is $1.40^{\circ}$, which is at least mostly due to the effects of
$D_e$ described above.  The magnitude of this effect will be quite
sensitive to the slope of the $D_e$ curve around $E_s=45$~eV.  We
therefore suspect that uncertainties in this slope are an important
source of systematic error in the {\em Ulysses} analysis.

     Could this effect in principle resolve the discrepancy in the
IBEX and {\em Ulysses} He flow vectors?  To investigate this
possibility, Figure~11 shows the location of the horseshoe track that
{\em Ulysses} should have observed if the IBEX vector of \citet{djm12}
was correct, approximated once again with the $T=0$~K
assumption.  In most places along the horseshoe, the $D_e$ slope
effect is not shifting the track towards the IBEX track, so assuming
$D_e$ curves with different slopes will unfortunately not resolve the
IBEX/{\em Ulysses} discrepancy.  The IBEX track in Figure~11 is an
average of $3.73^{\circ}$ from the observed track, illustrating the
magnitude of the disagreement in angular terms.  Based on Figure~7,
the lower $V_{ISM}$ and higher $\lambda$ values measured by IBEX (see
Section~1) would be expected to produce a horseshoe track that is
broader and shifted to the right from that observed by {\em Ulysses},
and Figure~11 shows that explicitly.

     Finally, we revisit the issue of He flow variability by
performing global fits to the {\em Ulysses} data separately for
the three fast latitude scans, leading to independent measurements
for 1994--1996, 2000--2002, and 2006--2007.  Results are listed
in Table~1 and displayed in Figure~12.  Consistent with the findings
of the empirical analysis of Section~3, we see no evidence for
any He flow parameter variability, with the He flow direction
varying by no more than $\sim 0.3^{\circ}$ from 1994--2007, and
the velocity varying by no more than $\sim 0.3$ km~s$^{-1}$.  With
IBEX data acquisition beginning in early 2009, there is less than
a two year time gap for the $3.5^{\circ}$ jump in flow longitude to
occur, if the  variable flow interpretation of the IBEX/{\em Ulysses}
discrepancy is correct \citep{pcf13}.


\section{DENSITY AND PHOTOIONIZATION}

     The previous section described how we measured four of the
five parameters describing the assumed Maxwellian He distribution in
the ISM.  The remaining parameter is the density, $n_{\rm He}$.  Unlike the
other four parameters, the density is significantly affected by loss
processes that occur during the He atoms' journey from the ISM to {\em
  Ulysses}.  Thus, measuring $n_{\rm He}$ is somewhat more complicated.  The
loss rate, $\beta_{ion}({\rm r})$, is assumed to vary as $r^{-2}$, and
is generally quantified by quoting the loss rate at 1~AU.  For He, the
most dominant loss process is photoionization by solar EUV photons,
though electron impact ionization may also be non-negligible
\citep{mw04,mb13}.  For an assumed $\beta_{ion}$ at 1~AU, our particle
tracking code can account for losses along the track, assuming the
$r^{-2}$ dependence.  At the ISM flow speed
it takes about 500 days to travel from 10 AU inward to the observer
(see, e.g., Figure~3 in Witte 2004), so the effective
ionization rate should be an average of the rates encountered over the
past year or so, weighted as $r^{-2}$.

     The usual procedure for dealing with photoionization is to
rely on estimates of $\beta_{ion}$ from various monitors of solar EUV
emission \citep{drm04,fa05,tnw05,jll11,mb13}.  However, this process
carries significant uncertainties, as many of the EUV irradiance
proxies are less than ideal for our purposes, and some of them are
inconsistent.  The Solar EUV Experiment (SEE) on the Thermospheric
Ionospheric Mesospheric Energy and Dynamics (TIMED) spacecraft
has significantly improved matters, but it was launched in 2002, and
so does not cover most of the {\em Ulysses} time period.  \citet{jll11}
quote uncertainties in EUV irradiance of 10\%-20\% in the SEE era,
but up to a factor of 2 uncertainties before it.  The SEE EUV
measurements have been correlated with other solar activity
indicators that have been monitored for many decades (e.g., Mg~II,
$F_{10.7}$), allowing those indicators to be utilized as EUV
proxies.  This has improved matters even for the older time period.

     Nevertheless, we here try a different technique for dealing
with photoionization that uses only {\em Ulysses} constraints.
Focusing on the second fast latitude scan (2000--2002), Figure~13a
shows densities measured as a function of orbital phase, computed
assuming different photoionization rates.  When an overly low
rate is assumed ($\beta_{ion}=0$ s$^{-1}$ in the figure) the result
is a U-shaped curve with low densities
at intermediate phases ($\phi=0.4-0.6$).  Although densities at
all phases will be underestimates, the effect is
more pronounced at these intermediate
phases, since that is when {\em Ulysses} is closest to the Sun and
therefore observes a He flow with the highest losses.
Conversely, when an overly high rate is assumed
($\beta_{ion}=3.0\times 10^{-7}$ s$^{-1}$ in the figure) the 
pattern reverses and overly high densities are observed at
intermediate phases, indicative of an overcompensation in the
loss correction.

     We define the most likely photoionization loss rate as that which
leads to the flattest density-vs-phase curve.  Specifically, we
compute the standard deviation of the density measurements divided by
the average density, $\sigma_n(\beta_{ion})/\bar{n}(\beta_{ion})$, and
we find the $\beta_{ion}$ value that minimizes this quantity.  At this
minimum, the average density $\bar{n}$ is then our $n_{\rm He}$ measurement,
and the minimum deviation is $\sigma_{n,min}/n_{\rm He}$.
We estimate 1$\sigma$ uncertainties for our loss rate and density
measurements using the standard deviation of
the mean at this minimum, $\sigma_{mean}=\sigma_{n,min}/\sqrt{N_m}$, where
here $N_m$ is the number of measurements.  The range of acceptable
loss rates are then defined to be where
\begin{equation}
\frac{\sigma_n(\beta_{ion})}{\bar{n}(\beta_{ion})}<
  \frac{\sigma_{n,min}+\sigma_{mean}}{n_{\rm He}}.
\end{equation}
For the epoch~2 data in Figure~13a, the minimum density variance is
at $\beta_{ion}=1.79\times 10^{-7}$ s$^{-1}$,
and the resulting densities are shown in the figure.  Assuming
this rate, the average He density is then computed to be
$n_{\rm He}=0.0197$ cm$^{-3}$.  Figure~13b displays the density
measurements for all three epochs.  The inferred densities and
photoionization rates are listed in Table~2, along with the
$1 \sigma$ uncertainties computed as described above.  For epoch~3,
the analysis is less constrained due to fewer data points and
intrinsically greater scatter (see Figure~13b), so we essentially
find only an upper limit to $\beta_{ion}$.

     The densities independently measured from the three epochs are
not identical.  Assuming that the ISM He density does not really vary,
the degree of inconsistency is presumably an indication of the
uncertainties in the loss correction, although detector sensitivity
uncertainties are also a potential factor.  The stability of the
detector sensitivity was tested twice during the course of the
mission.  New layers of LiF were evaporated onto the surface from the
GAS instrument's LiF supply in 1995 and 2001. The He signal was
measured immediately before and after each deposition. It showed that
the detection efficiency did not change
within the available statistical accuracy of 20\%.  Unfortunately, a
third test planned for 2006 had to be cancelled, as the power
necessary to heat the LiF supply may have exceeded the power
available to the spacecraft at the time, which had decreased during
the course of the mission.  Thus, it cannot be proven that the
detection efficency did not degrade during the third and final epoch,
altough it seems unlikely in view of past experience.  In any case, the
average and standard deviation of the three measurements in Table~2
provides our final best estimate for the ISM He density:
$n_{\rm He}=0.0196\pm 0.0033$ cm$^{-3}$.  This is somewhat higher than
the $n_{\rm He}=0.015\pm 0.003$ cm$^{-3}$ measurement from \citet{mw04},
due at least in part to higher assumed photoionization rates, but
the error bars overlap.

     Table~2 compares our inferred loss rates with average ones
estimated by \citet{mb13,mb14} for the three epochs based on
solar wind and EUV constraints, taking into consideration both
photoionization and electron impact ionization.  There is actually
good agreement for epochs 2 and 3, but not for epoch 1, where our
$\beta_{ion}=(1.36\pm 0.27)\times 10^{-7}$ s$^{-1}$ is significantly higher
than the $\beta_{ion}=0.9\times 10^{-7}$ s$^{-1}$ rate expected
for this period.  Given that our estimated density is also
rather high, it seems likely that our analysis approach has
led to an overestimate of $\beta_{ion}$.  Our technique of
minimizing phase-dependent density variation does have the
drawback of not accounting for any change in photoionization
rate that might occur in the 20 month period covering the
$\phi=0.25-0.90$ fast-latitude scan range.  However, the loss rates
estimated by \citet{mb13,mb14} do not actually show much variation
during epoch 1, so unidentified issues may also be present.

\section{SUMMARY}

     Motivated in part by the discrepancy in the He flow
vectors measured by {\em Ulysses} and IBEX, we have performed
an independent analysis of the {\em Ulysses}/GAS data set
from 1994-2007, newly reprocessed in 2013, including 2006--2007
data not included in the canonical \citet{mw04} analysis.  Our
findings are summarized as follows:
\begin{description}
\item[1.] We first perform a purely empirical analysis of the
  He beam observed by {\em Ulysses}, using 2-D Gaussian fits to
  quantify beam location, amplitude, and width; in order to see
  how these parameters vary with orbital phase and time.  There is
  no evidence for intrinsic variability in beam direction by more
  than a few tenths of a degree.  The beam widths are also invariant
  within measurement uncertainties.
\item[2.] The lack of variability in the He beam
  from 1994--2007 makes it harder to explain the
  IBEX/{\em Ulysses} discrepancy in terms of He flow
  variability, with less than a two-year time gap for that change
  in flow to occur before IBEX observations begin in 2009.
\item[3.] We demonstrate that the most powerful constraint that
  the {\em Ulysses}/GAS data provide on the He flow vector
  lies in the shape of the horseshoe-shaped track made by the He
  beam over the course of a {\em Ulysses} orbit,
  which breaks parameter degeneracies that plague attempts to
  derive He flow parameters from individual beam maps.  Only
  by fitting the {\em Ulysses}/GAS data collectively can this
  diagnostic power be fully utilized.
\item[4.] From a global fit to the full {\em Ulysses}/GAS data set
  from 1994--2007, we find the following He flow parameters:
  $V_{ISM}=26.08\pm 0.21$ km~s$^{-1}$, $\lambda=75.54\pm 0.19^{\circ}$,
  $\beta=-5.44\pm 0.24^{\circ}$, and $T=7260\pm 270$~K.  Separate
  global fits to the three separate fast-latitude scans confirm
  a lack of He flow parameter variability, with the flow direction
  varying by no more than $\sim 0.3^{\circ}$ and the velocity by
  no more than $\sim 0.3$ km~s$^{-1}$.
\item[5.] Our ($V_{ISM}$,$\lambda$,$\beta$) He vector is consistent
  with the previous canonical measurements of \citet{mw04} and the
  contemporaneous analysis of \citet{mb14}, proving
  that the IBEX/{\em Ulysses} discrepancy is not a product of
  different analysis approaches.  The cause of the
  IBEX/{\em Ulysses} discrepancy remains unknown.
\item[6.] Our temperature measurement ($T=7260\pm 270$~K) is somewhat
  higher than the $T=6300\pm 340$~K value of \citet{mw04}, but is
  consistent with the $T=7500$~K best-fit value of \citet{mb14},
  although \citet{mb14} quote a much larger range of acceptable
  values ($5500<T<9000$~K).
  A higher temperature potentially mitigates the
  IBEX/{\em Ulysses} discrepancy somewhat, as the parameter
  dependencies in the IBEX analysis are such that a higher
  $T$ would lead to lower $\lambda$ and higher $V_{ISM}$, moving
  the IBEX-derived values closer to those of {\em Ulysses}.
\item[7.] Using a technique to infer photoionization loss rates
  solely from {\em Ulysses} constraints, we compute He densities and
  loss rates for the three epochs of observation (1994--1996, 2000--2002,
  and 2006--2007).  Our loss rates agree well with expectations from
  solar EUV constraints for epochs 2 and 3, but not for epoch 1,
  where we seem to overestimate the rate.  Our final
  density measurement is $n_{\rm He}=0.0196\pm 0.0033$ cm$^{-3}$,
  somewhat higher than estimated by \citet{mw04},
  $n_{\rm He}=0.015\pm 0.003$ cm$^{-3}$.
\end{description}

\acknowledgments

We would like to thank Dr.\ E.\ M\"{o}bius for useful discussion and
suggestions.  This work has been supported by NASA award NNH13AV19I to
the Naval Research Laboratory.

\clearpage

\begin{deluxetable}{ccccccc}
\tabletypesize{\scriptsize}
\tablecaption{{\em Ulysses} Measurements of the ISM He Flow}
\tablecolumns{7}
\tablewidth{0pt}
\tablehead{
  \colhead{Years} & \colhead{Fit Type} & \colhead{$V_{ISM}$}&
    \colhead{$\lambda$ (J2000)} & \colhead{$\beta$ (J2000)} &
    \colhead{T} & \colhead{Source} \\
  \colhead{}       & \colhead{}         & \colhead{(km s$^{-1}$)} &
     \colhead{(deg)} & \colhead{(deg)} & \colhead{(K)} & \colhead{}}
\startdata
1994--2007    & Global Fit & $26.08\pm 0.21$ & $75.54\pm 0.19$ &
  $-5.44\pm 0.24$ & $7260\pm 270$ & this work \\
1994--2007    &Single Maps & (26.08)         & $75.60\pm 0.34$ &
  $-5.38\pm 0.37$ & $7090\pm 370$ & this work \\
1994--1996    & Global Fit & $26.14\pm 0.24$ & $75.53\pm 0.25$ &
  $-5.50\pm 0.29$ & $7230\pm 360$ & this work \\
2000--2002    & Global Fit & $26.08\pm 0.23$ & $75.57\pm 0.22$ &
  $-5.39\pm 0.27$ & $7230\pm 310$ & this work \\
2006--2007    & Global Fit & $25.92\pm 0.25$ & $75.42\pm 0.17$ &
  $-5.30\pm 0.30$ & $7270\pm 220$ & this work \\
1990--2002    &Single Maps & $26.3\pm 0.4$ & $75.4\pm 0.5$ &
  $-5.2\pm 0.2$   & $6300\pm 340$ & Witte (2004)\\
\enddata
\end{deluxetable}

\begin{deluxetable}{ccccccc}
\tabletypesize{\normalsize}
\tablecaption{Density and Photoionization}
\tablecolumns{7}
\tablewidth{0pt}
\tablehead{
  \colhead{Years} & \colhead{$n_{\rm He}$ from Fit} & \colhead{$\beta_{ion}$ from Fit} &
    \colhead{$\beta_{ion}$ from EUV\tablenotemark{a}} \\
  \colhead{}  & \colhead{($10^{-2}$ cm$^{-3}$)} & \colhead{($10^{-7}$ s$^{-1}$)} &
     \colhead{($10^{-7}$ s$^{-1}$)}}
\startdata
1994--1996 & $2.28\pm 0.14$     & $1.36\pm 0.27$      & 0.9 \\
2000--2002 & $1.97\pm 0.18$     & $1.79\pm 0.31$      & 1.6 \\
2006--2007 & $1.62^{+0.58}_{-0.31}$ & $0.57^{+0.92}_{-0.57}$ & 0.7 \\
\enddata
\tablenotetext{a}{From Bzowski et al.\ (2013).}
\end{deluxetable}

\clearpage

\begin{figure}[t]
\plotfiddle{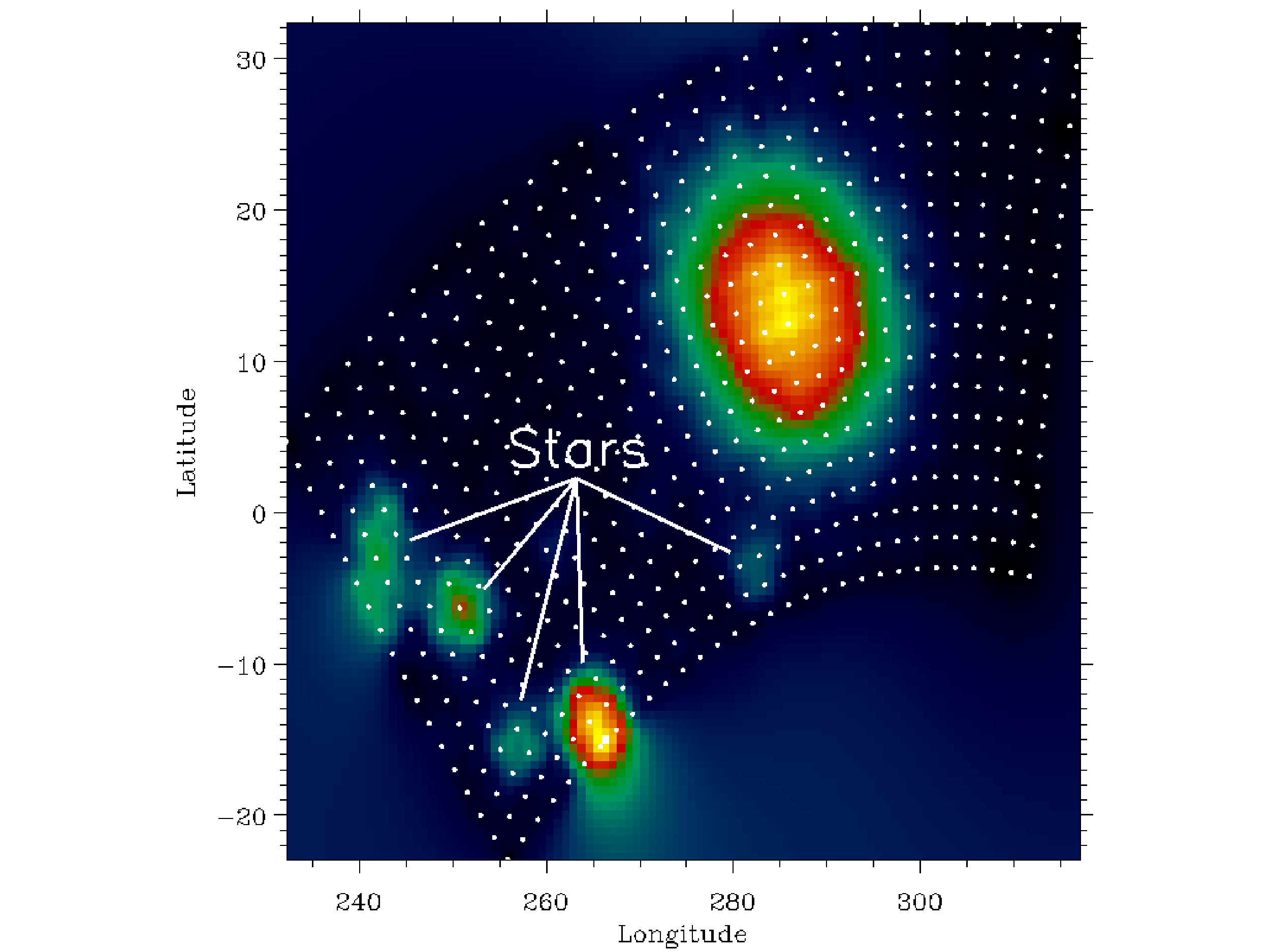}{3.5in}{0}{80}{80}{-290}{-10}
\caption{A WFOV helium beam map made by {\em Ulysses}/GAS on
  2001~January~24, in ecliptic coordinates.  White dots indicate
  actual scan positions used to make the map.  In addition to the
  broad He beam, several stars are also visible.}
\end{figure}

\clearpage

\begin{figure}[t]
\plotfiddle{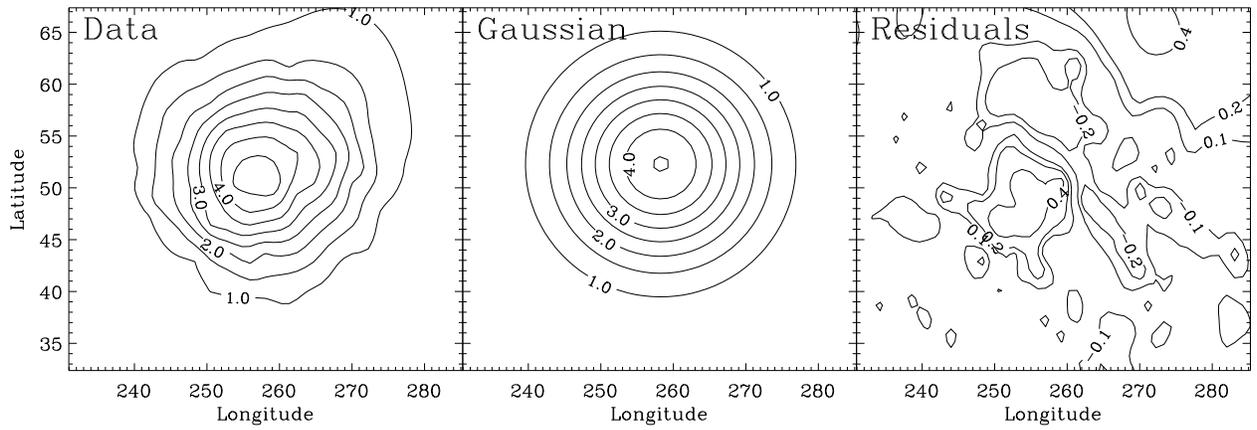}{1.5in}{0}{95}{95}{-290}{-490}
\caption{The left panel is a WFOV helium beam map from 2001~July~6,
  in ecliptic coordinates, the middle panel is a 2-D Gaussian fit to
  the He peak, and the right panel shows the residuals of the fit.
  The count rates shown are in cts~s$^{-1}$~pixel$^{-1}$.}
\end{figure}

\clearpage

\begin{figure}[t]
\plotfiddle{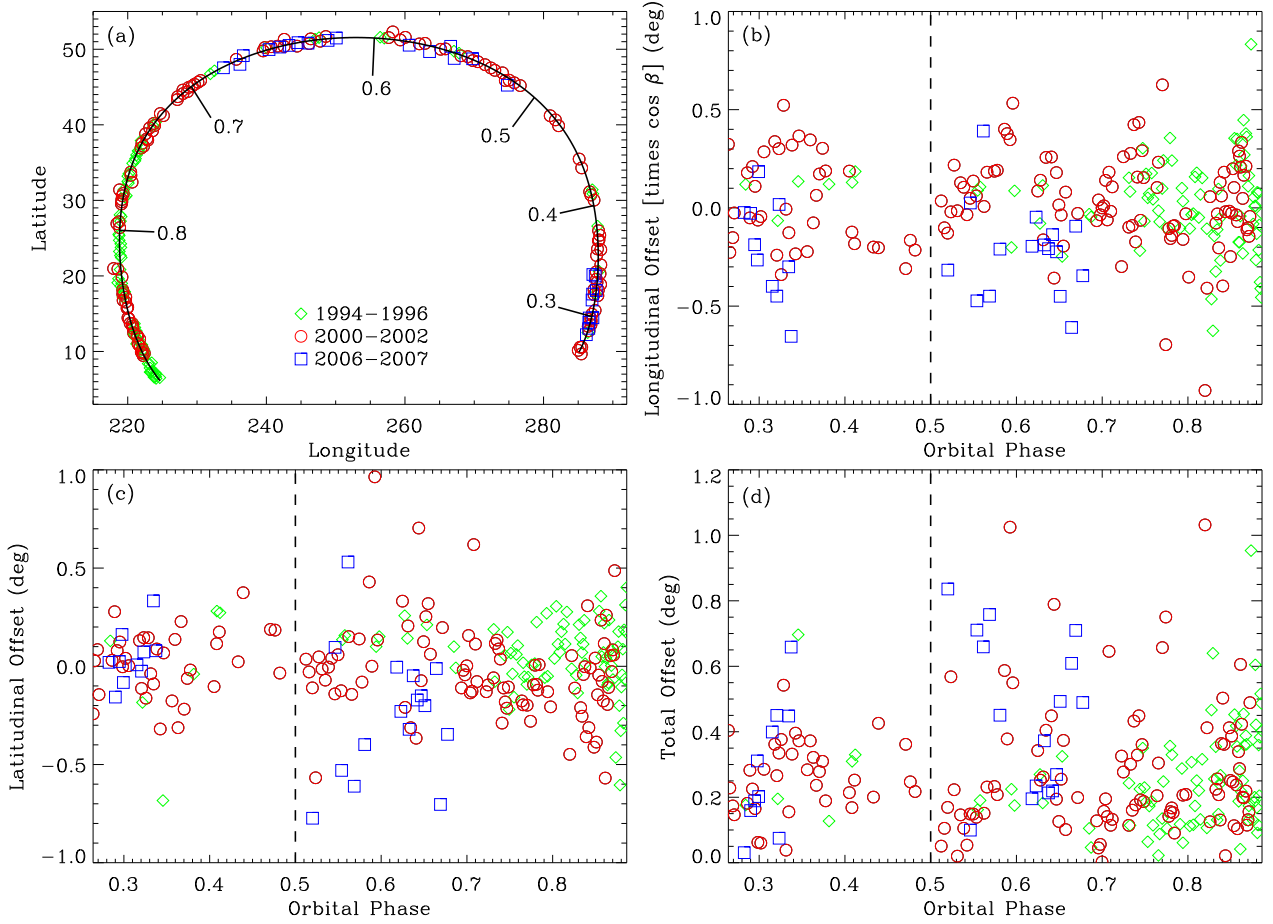}{4.0in}{90}{70}{70}{280}{-50}
\caption{(a) The center of the helium beam as measured by the 2-D
  Gaussian fits, in ecliptic coordinates, with different symbols
  indicating measurements from the three different fast latitude
  scans.  During the course of a fast latitude scan the He beam traces
  out this horseshoe pattern in the sky.  The solid line is a
  polynomial fit to this horseshoe.  The numbers along the horseshoe
  indicate orbital phase, where an orbital phase of 0.5 corresponds to
  the ecliptic plane crossing near perihelion.  (b) Longitudinal
  discrepancies from the average beam location shown in (a), with the
  dashed line indicating the ecliptic plane crossing.  (c)
  Latitudinal discrepancies from the average beam location.  (d) Total
  angular offset from the average beam location.}
\end{figure}

\clearpage

\begin{figure}[t]
\plotfiddle{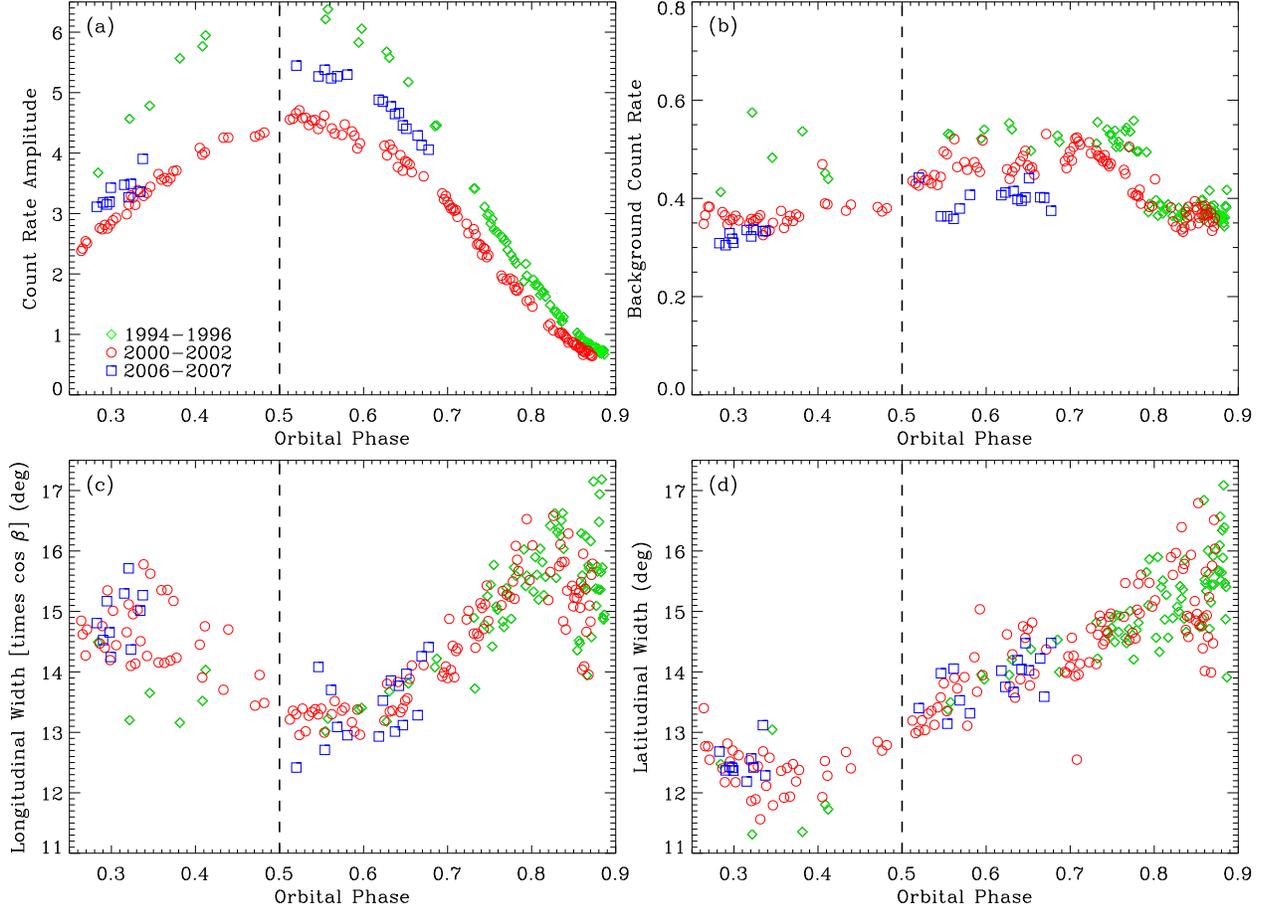}{4.0in}{90}{70}{70}{280}{-50}
\caption{(a) Peak count rate amplitude of the helium beam based on the
  2-D Gaussian fits, as a function of orbital phase, where an orbital
  phase of 0.5 corresponds to the ecliptic plane crossing near
  perihelion (dashed line).  Different symbols indicate measurements
  from the three different fast latitude scans.  The count rates shown
  are in cts~s$^{-1}$~pixel$^{-1}$.  The other three panels show (b)
  background count rates, (c) longitudinal beam widths, and (d)
  latitudinal beam widths.}
\end{figure}

\clearpage

\begin{figure}[t]
\plotfiddle{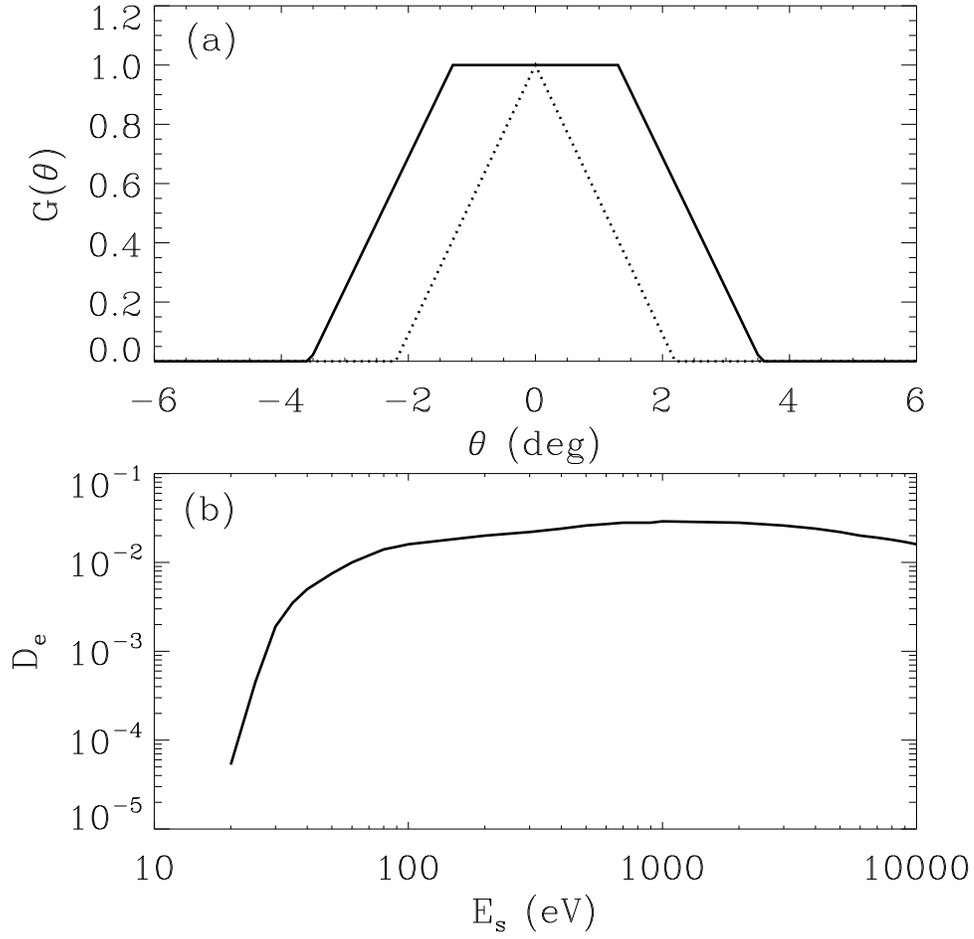}{3.0in}{90}{70}{70}{280}{-40}
\caption{(a) Detector geometry function for the WFOV (solid line) and
  NFOV (dotted line) detectors.  (b) Detection efficiency
  as a function of incoming particle energy.}
\end{figure}

\clearpage

\begin{figure}[t]
\plotfiddle{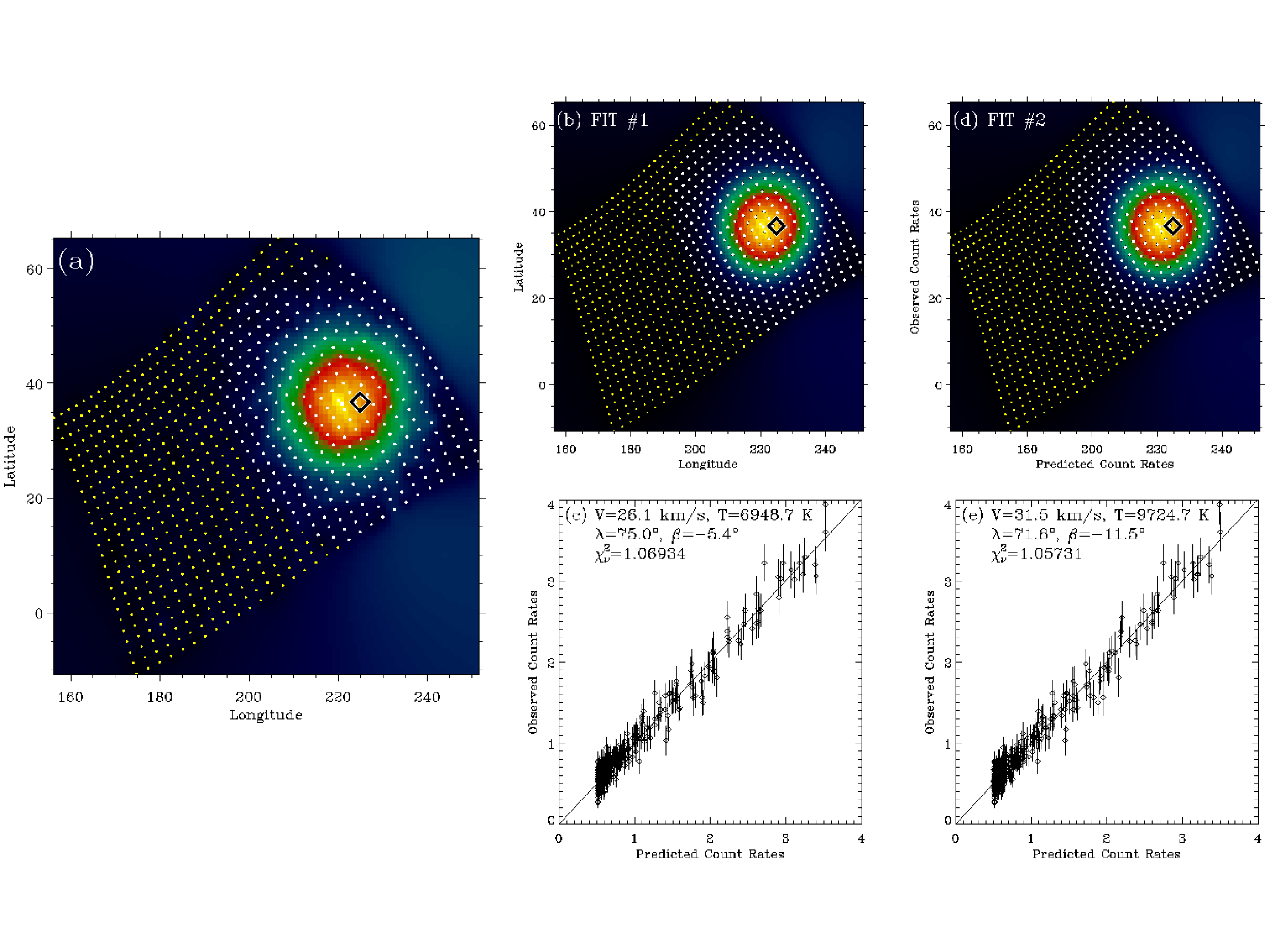}{3.5in}{0}{68}{68}{-250}{-30}
\caption{(a) A WFOV helium beam map from 1995~August~20, in ecliptic
  coordinates.  The diamond indicates the expected beam position based
  on the Witte (2004) vector (assuming $T=0$~K).  Yellow dots are scan
  positions, and white dots are the measurements within $25^{\circ}$
  of the diamond, which are the data considered in fits.  (b) A
  synthetic image representing a fit to the image (FIT \#1) in (a),
  where the Witte (2004) vector was the initial guess for the best-fit
  parameters.  (c) Observed versus predicted count rates for the fit
  from (b).  Parameters of this fit are listed, as well as the reduced
  $\chi^2$ value ($\chi^2_{\nu}=1.06934$).  (d) Another fit (FIT \#2)
  to the data with a very different initial guess for the fit
  parameters from (b).  (e) Analogous to (c) but for FIT \#2.  Note
  the very different best-fit parameters compared to (c).}
\end{figure}

\clearpage

\begin{figure}[t]
\plotfiddle{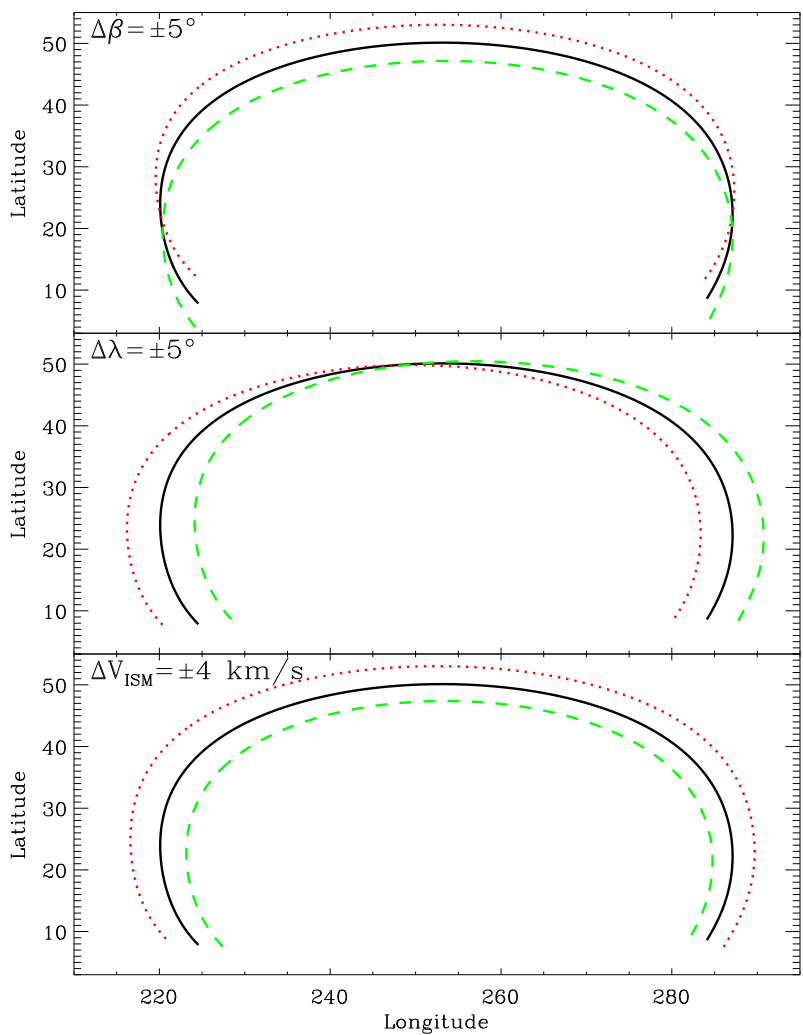}{4.0in}{90}{80}{80}{310}{-50}
\caption{The solid line in each panel indicates the variation of He
  beam location with orbital phase expected from the Witte (2004)
  flow vector.  Also shown in each panel is what happens to this
  horseshoe-shaped beam track when the latitude ($\beta$), longitude
  ($\lambda$), or velocity ($V_{ISM}$) of the vector are increased
  (dashed lines) or decreased (dotted lines) by the amount indicated.}
\end{figure}

\clearpage

\begin{figure}[t]
\plotfiddle{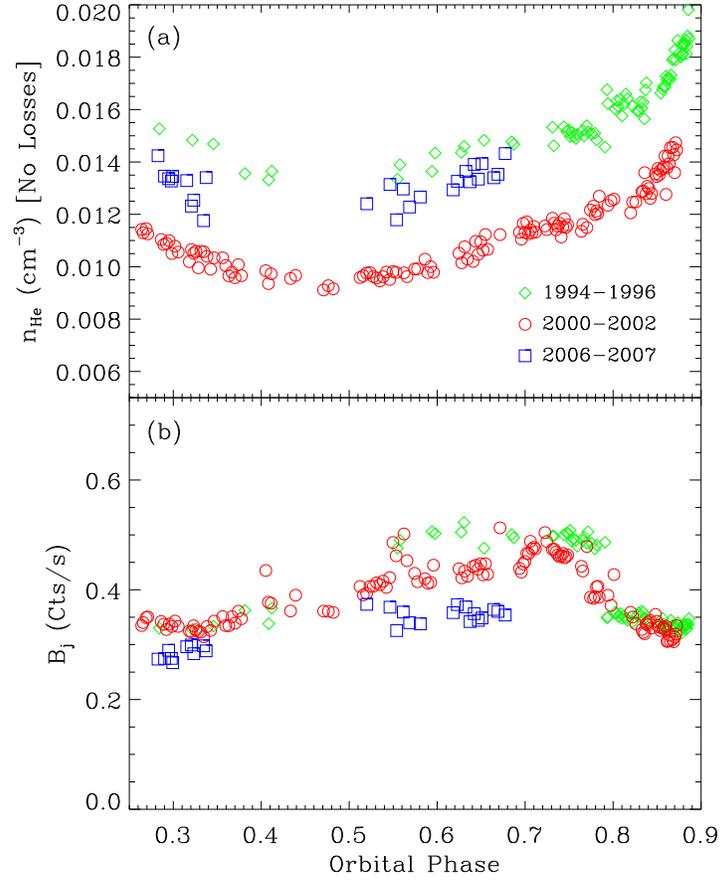}{3.0in}{90}{70}{70}{280}{-40}
\caption{(a) Inferred He densities from a global fit to 238 He beam
  maps, with no correction for photoionization losses.  The phase
  dependence is indicative of the presence of such losses.  Different
  symbols are for different epochs, as in Figures~3--4. (b) Background
  count rates measured for the 238 maps in the global fit.}
\end{figure}

\clearpage

\begin{figure}[t]
\plotfiddle{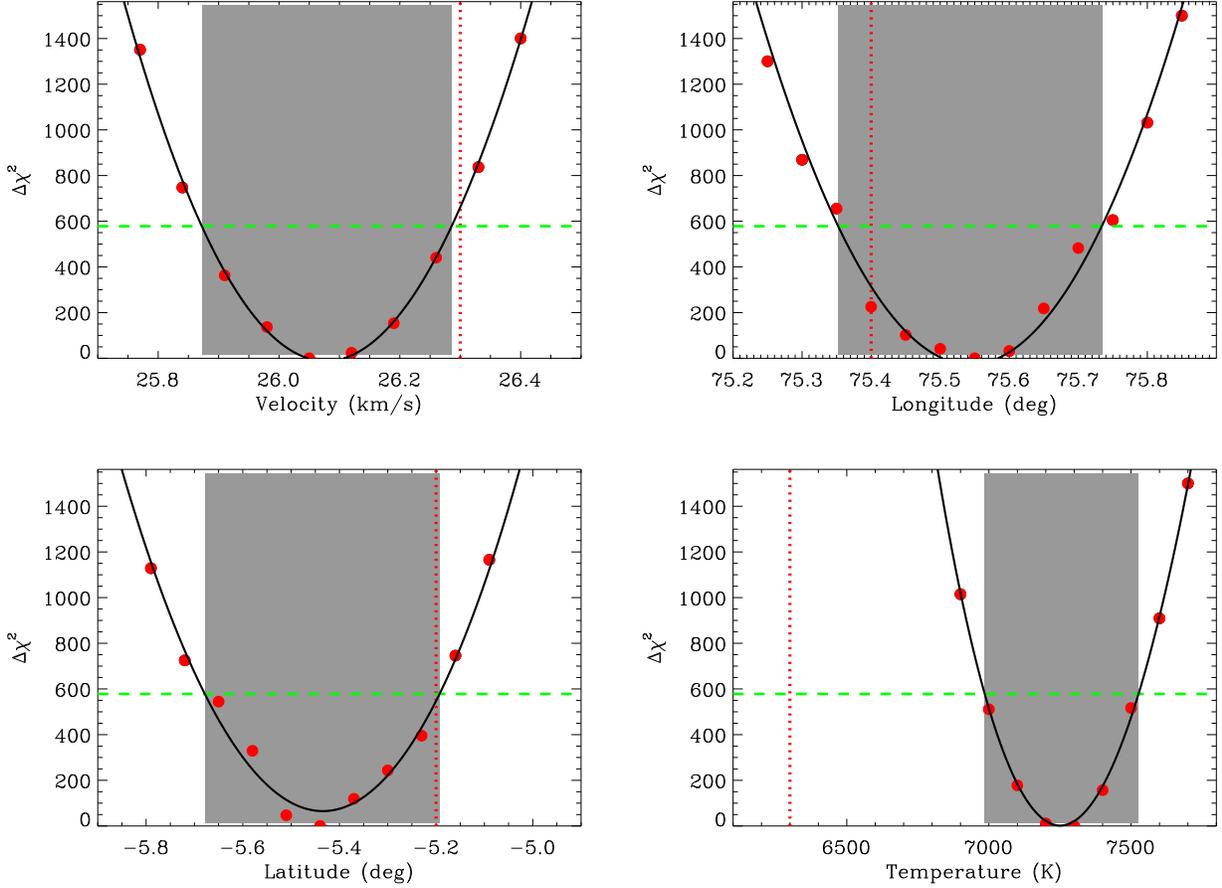}{3.0in}{90}{70}{70}{275}{-50}
\caption{Helium flow parameter measurements from a global fit to the
  {\em Ulysses}/GAS data.  In each panel, we plot $\Delta \chi^2$ as a
  function of one of the four He flow parameters of interest, where
  each point represents a separate fit with that parameter held
  constant and the other three parameters allowed to vary freely.
  Solid lines show polynomial fits to the data points.  The horizontal
  dashed line corresponds to a $3\sigma$ contour, which we use to
  define our uncertainty range in each parameter (shaded regions).
  Vertical dotted lines are the Witte (2004) results, which are only
  significantly discrepant for the temperature.}
\end{figure}

\clearpage

\begin{figure}[t]
\plotfiddle{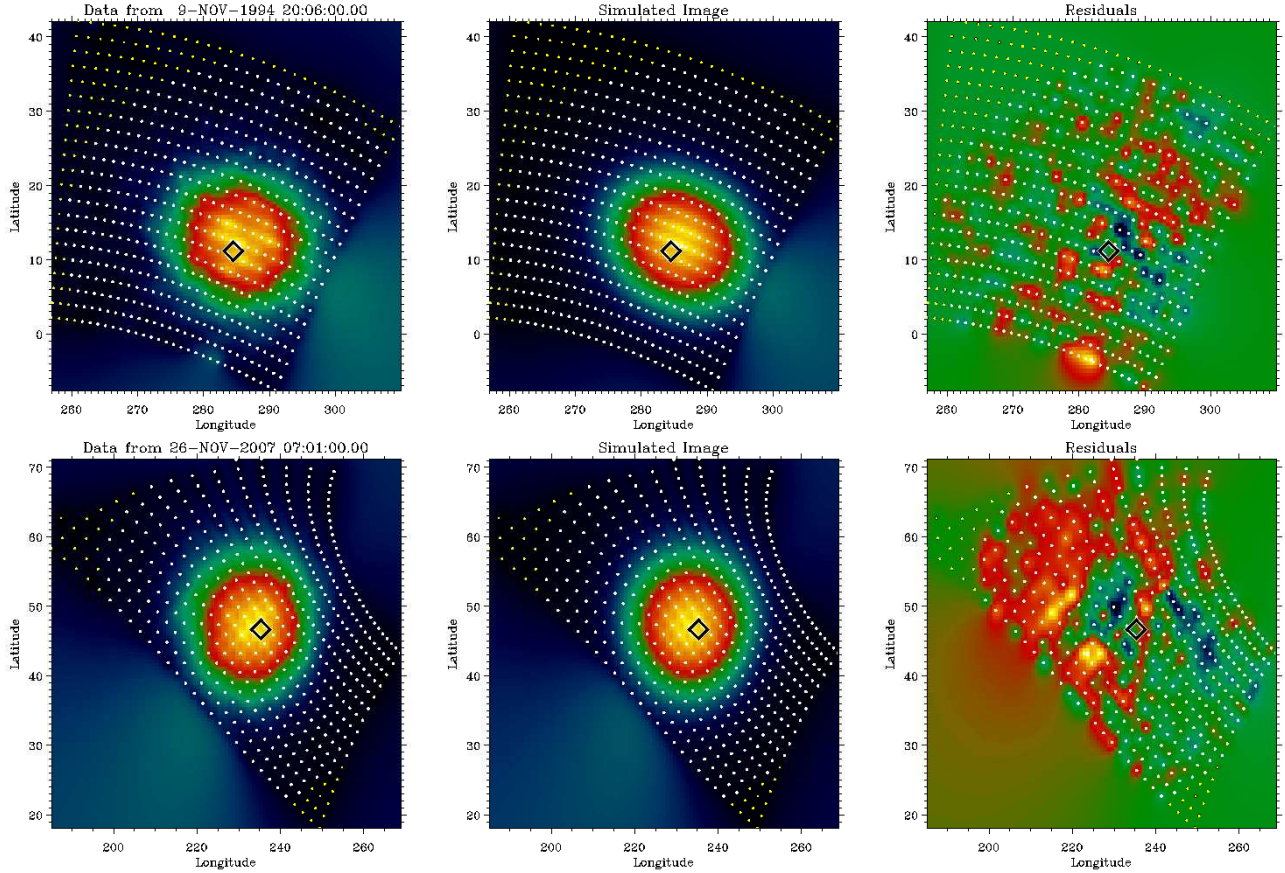}{3.5in}{0}{68}{68}{-270}{-40}
\caption{The top row shows a He beam map from 1994~November~9, a simulated
  map based on our best global fit to the {\em Ulysses} data, and the
  residuals of this fit.  The bottom row is a similar set of images
  for a He beam map from 2007~November~26.  The diamond indicates the
  expected beam position based on the Witte (2004) vector (assuming
  $T=0$~K).  Yellow dots are scan positions, and white dots are the
  measurements within $25^{\circ}$ of the diamond, which are the data
  considered in the fit.}
\end{figure}

\clearpage

\begin{figure}[t]
\plotfiddle{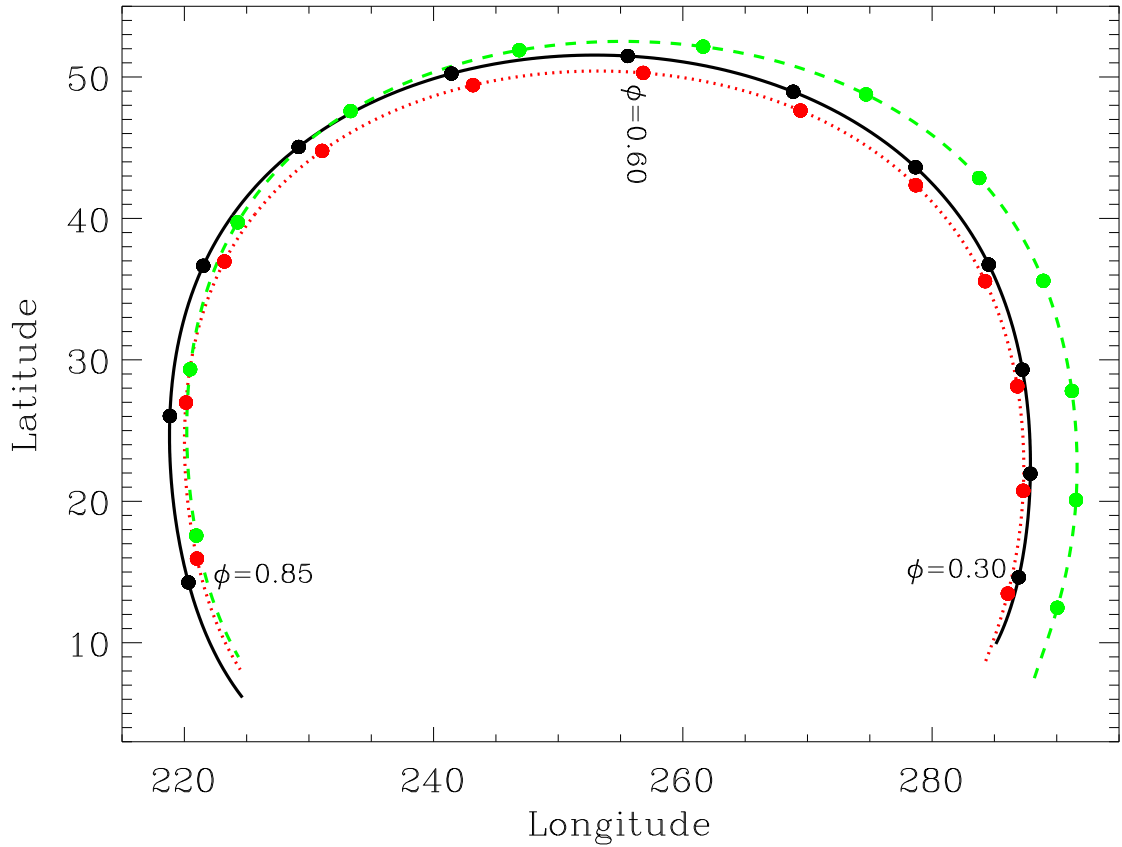}{3.0in}{90}{70}{70}{270}{-50}
\caption{The solid line is the observed track of the He beam center
  during the {\em Ulysses} orbit, from Figure~3a, which is compared
  with the expected beam center based on our best-fit He flow
  vector (dotted line) and the IBEX-derived flow vector (dashed line)
  from McComas et al.\ (2012).  The expected beam centers are computed
  assuming $T=0$~K, and the difference between the best-fit and observed
  tracks indicates the level of uncertainty introduced by that assumption
  (see text).  Dots mark orbital phases between
  $\phi=0.30$ and $\phi=0.85$ in increments of $\Delta \phi=0.05$.}
\end{figure}

\clearpage

\begin{figure}[t]
\plotfiddle{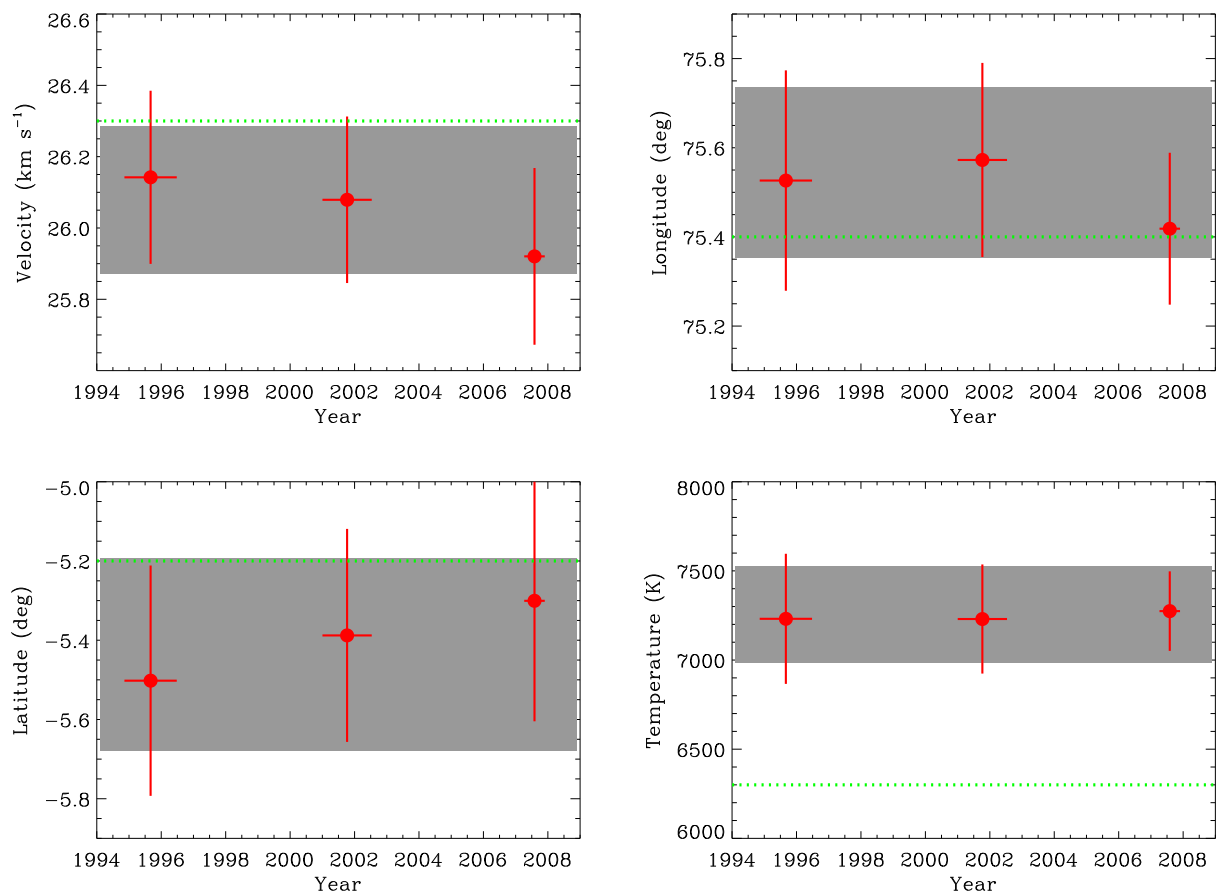}{3.0in}{90}{70}{70}{275}{-50}
\caption{Helium flow parameter measurements from global fits to
  {\em Ulysses}/GAS data from the three fast latitude scans in 1994--1996,
  2000--2002, and 2006--2007.  The shaded regions correspond to
  the error bars defined in Figure~9.  The horizontal dotted lines
  are the Witte (2004) parameters.  We see no evidence for any
  He flow parameter variability from 1994--2007.}
\end{figure}

\clearpage

\begin{figure}[t]
\plotfiddle{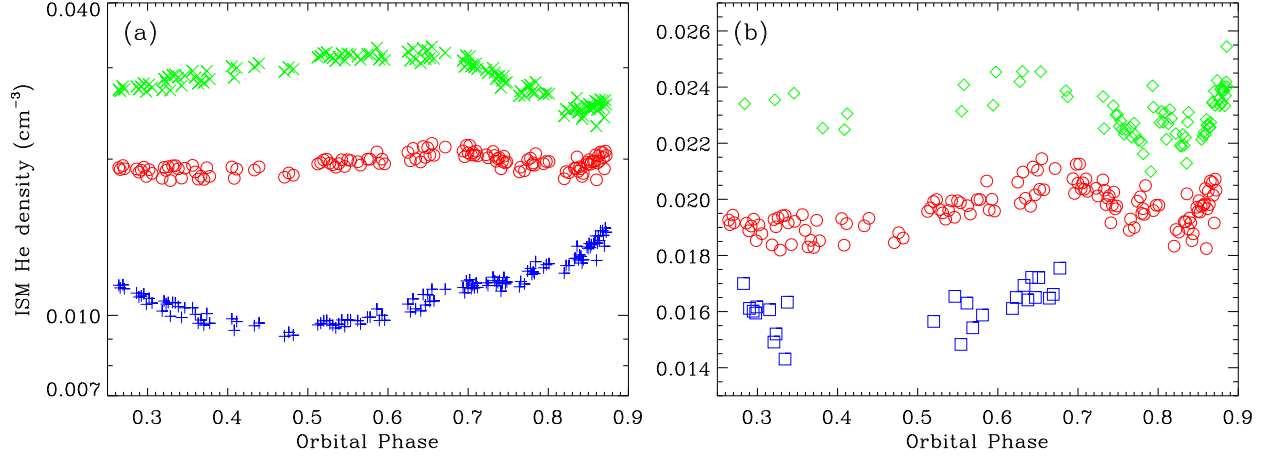}{0in}{90}{70}{70}{280}{-210}
\caption{(a) Interstellar He densities computed from the second
  fast latitude scan (2000--2002), plotted versus orbital phase,
  assuming photoionization loss rates of $\beta_{ion}=0$ s$^{-1}$ (plus signs),
  $1.79\times 10^{-7}$ s$^{-1}$ (circles), and $3.0\times 10^{-7}$ (X's) s$^{-1}$.
  The $\beta_{ion}=1.79\times 10^{-7}$ s$^{-1}$ case is where the
  variability of the densities is minimized.  (b) Inferred He
  densities plotted as a function of orbital phase for the
  first (diamonds), second (circles), and third (squares) fast
  latitude scans.  For each of the three sets of measurements the
  loss rate is chosen to minimize variability in density with
  orbital phase.  These rates are $1.36\times 10^{-7}$,
  $1.79\times 10^{-7}$, and $5.7\times 10^{-8}$ s$^{-1}$
  for scans one, two, and three, respectively.}
\end{figure}

\end{document}